# Properties of Confined Ammonium Nitrate Ionic Liquids

Andrei Filippov, *,a,b and Oleg N. Antzutkin[a,c]

[a]Chemistry of Interfaces, Luleå University of Technology, SE-97187 Luleå, Sweden
[b]Institute of Physics, Kazan Federal University, 420008 Kazan, Russia
[c]Department of Physics, Warwick University, Coventry CV4 7AL, UK

Andrei.Filippov@ltu.se

Ethylammonium nitrate (EAN) and propylammonium nitrate ionic liquids confined between polar glass plates and exposed to a strong magnetic field demonstrate gradually slowing diffusivity, a process that can be reversed by removing the sample from the magnetic field. The process can be described well by the Avrami equation, which is typical for autocatalytic (particularly, nucleation controlled) processes. The transition can be stopped by freezing the sample. Cooling and heating investigations showed differences in the freezing and melting behavior of the sample depending on whether it had been exposed to the magnetic field. After exposure to the magnetic field, the sample demonstrated a change in the state of residual water. Generally, our findings confirm our previous suggestion that alteration of the dynamic properties of confined ammonium nitrate ionic liquids exposed to a magnetic field is related to the alteration of real physical-chemical phases.

## 1. Introduction

Ionic liquids (ILs) are prepared from organic cations and either organic or inorganic anions.[1,2] They have been used as electrolyte materials in lithium batteries[3,4] and ultracapacitors,[5] as media for chemical reactions, protein separation,[2,6] $CO_2$ absorption,[7] and as lubricants.[8] Ammonium nitrates have three readily exchangable protons on the $NH_3$ group of cations, therefore, they belong to a class of so called 'protic' ionic liquids.[1] EAN is the most frequently reported protic IL,[1] which is used as a medium for chemical reactions, as a precipitating agent for protein separation,[6] and as an electrically conductive solvent in electrochemistry.[3] Like water, EAN has a three-dimensional hydrogen-bonding network and can be used as an amphiphilic self-assembly medium.[9] Confined ILs,[10-12] particularly EAN,[11,13,14] have attracted special interest during the last few years. Enhanced diffusion of ethylammonium (EA) cations has been observed for EAN confined between polar glass plates.[13] It was experimentally found that both self-diffusion and NMR relaxation of EAN confined between polar glass plates reversibly alter after placement in a strong magnetic field.[14] The main factors responsible for this effect are the availability of protons in the protic EAN and the polarity of the surface, while the exchange rate of -$NH_3$ protons plays a crucial role in the observed processes. It has been suggested that the processes influencing the dynamics of EAN in this confinement are the phase transformations of EAN.[14]

In this work, we further investigate the physical properties of confined EAN and another protic IL belonging to the same class, propylammonium nitrate (PAN), and the effect of applying a strong magnetic field under the same confinement conditions, i. e. between polar glass plates.

## 2. Materials and methods

### 2.1. Sample preparation

The chemical structures of ions of the studied ionic liquids are shown in Figure 1. ILs were synthesized and characterized at the Chemistry of Interfaces of Luleå University of Technology as described previously.[13] EAN and PAN are liquids at ambient conditions. The quantity of water in the synthesized ILs was less than 0.55 % as determined by Karl-Fisher titration (Metrohm 917 Karl Fisher Coloumeter with HYDRANAL reagent).

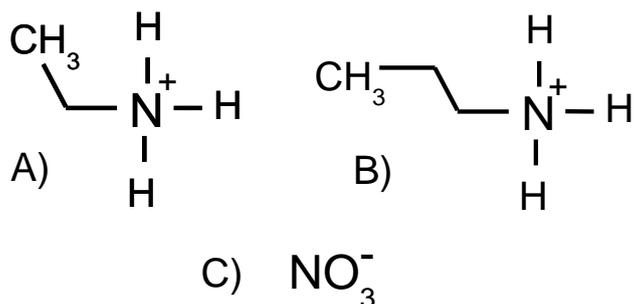

**Figure 1.** Chemical structures of: **A**) ethylammonium (EA) cation; **B**) propylammonium cation; and **C**) nitrate anion.

NMR measurements for the bulk ionic liquids were performed by placing 300 µl of the IL in a standard 5-mm NMR glass tube. Confined ILs were prepared with glass plates arranged in a stack. The plates (14 x 2.5 x 0.1 mm, Thermo Scientific Menzel-Gläser, Menzel GmbH, Germany) were carefully cleaned before sample preparation (see the ESI). Contact angle measurements with Milli-Q water gave a contact angle near 0°. A stack of glass plates filled with IL was prepared in a glove box in a dry $N_2$ atmosphere. Samples were made by adding 2 µL of an IL to the first glass plate, placing a new glass plate on top, adding 2 µL of the IL on top of this glass plate, etc. until the thickness of the stack reached 2.5 mm. IL that overflowed at the edges of the stack was removed by sponging. The sample consisting of a stack of *ca.* 37 glass plates was placed in a rectangular glass tube and sealed. The mean spacing between the glass plates was assessed by weighing the introduced IL, which yielded $d$ ~3.8-4.5 µm for EAN.[13] A detailed report of the sample preparation and characterization has been described in our previous papers.[13,14] To allow equilibration of ILs inside the samples, experiments were started a week after the sample preparation.

### 2.2. Pulsed-field gradient diffusometry

$^1$H NMR self-diffusion measurements were performed using a Bruker Ascend/Aeon WB 400 (Bruker BioSpin AG, Fällanden, Switzerland) NMR spectrometer with a working frequency of 400.27 MHz for $^1$H, magnetic field strength of 9.4 T with magnetic field homogeneity better than $4.9 \cdot 10^{-7}$ T, using a Diff50 Pulsed-Field-Gradient (PFG) probe. An NMR solenoid $^1$H insert was used to macroscopically align the plates of the sample stack at 0 and 90 degrees with respect to the direction of the external magnetic field (the same direction as that of the PFG).

The BCU II cooling unit was used for low-temperature NMR experiments. This allows maintenance of the temperature of the NMR sample at temperatures down to approximately 230 K with precision of 1 K. The unit was calibrated using copper-constantan thermocouple.

The diffusional decays (DD) were recorded using the spin-echo (SE, at $t_d \leq 5$ ms) or the stimulated echo (StE) pulse sequences. For single-component diffusion, the form of DD can be described as follows:[15,16]

$$A(\tau, g, \delta) \propto \exp\left(-\frac{2\tau}{T_2}\right) \exp\left(-\gamma^2 \delta^2 g^2 D t_d\right) \quad (1a)$$

$$A(\tau, \tau_1, g, \delta) \propto \exp\left(-\frac{2\tau}{T_2} - \frac{\tau_1}{T_1}\right) \exp\left(-\gamma^2 \delta^2 g^2 D t_d\right) \quad (1b)$$

for the SE and StE, respectively. Here, $A$ is the signal intensity, $\tau$ and $\tau_1$ are the time intervals in the pulse sequence; $\gamma$ is the gyromagnetic ratio for protons; $g$ and $\delta$ are the amplitude and the duration of the gradient pulse; $t_d = (\Delta - \delta/3)$ is the diffusion time; $\Delta = (\tau + \tau_1)$. In the measurements, the duration of the 90° pulse was 7 μs, $\delta$ was in the range of 0.5-2 ms, $\tau$ was in the range of 3-5 ms, and the amplitude of $g$ was varied from 0.06 up to 29.73 T·m$^{-1}$. The recycle delay was 3.5 s.

As has been shown,[13] restrictions (glass plates) have no effect on the $D$ of the IL in the direction along the plates, while for diffusion normal to the plates the effect of the plates on $D$ was perceptible at $t_d$ longer than 3 ms. For diffusion in bulk and along the plates, $D$ values were acquired by fitting Eq. (1) to the experimental decays. For decays obtained in the direction normal to the plates at $t_d = 3$ ms, $D$ was calculated from the equation:

$$D_{av} = \frac{-\partial A(\gamma^2 \delta^2 g^2 t_d)}{\partial (\gamma^2 \delta^2 g^2 t_d)}\bigg|_{(\gamma^2 \delta^2 g^2 t_d) \to 0} \quad (2)$$

Most of measurements were performed at 293 K (room temperature), therefore, we did not wait until thermal equilibration of the sample in the NMR probe was reached and measurements were started directly (in 1-2 min) after placing the sample in the NMR probe, tuning and matching the probe. The time required for a single NMR diffusion measurement was around 30 s. Data were processed using Bruker Topspin 3.5 software.

## 3. Results and discussion

In previous[13,14] and current studies we observed differences in self-diffusion and local dynamics of ammonium nitrates (mainly EAN) in their (i) bulk states, (ii) confined between polar glass plates and (iii) the system (ii) exposed for rather long time (*ca.* 24 h) in a strong static magnetic field. These states were characterized as "dynamic phases".[13,14] For the purpose of simplicity, in further discussion we use a term "phase" for ammonium nitrates in the states (i), (ii) and (iii) as **I$_B$**, **I** and **I$_{MF}$**, respectively, taking into account that we have not yet proven structural and thermodynamic differences between phases of EAN in these states.

### 3.1. Diffusivity
Placement of bulk ILs in the magnetic field of the NMR spectrometer and systematic measurements of $D$ during and over 24 hours showed no changes in diffusivity.

Therefore, it can be suggested that the bulk phases "$I_B$" of this class of protic ILs were maintained during magnetic field exposure according to previous findings.[13,17] The values of the diffusion coefficients ($D_0$) for bulk EAN and PAN are shown in Figure 2 as dotted lines. The diffusion coefficients of cations of EAN and PAN between polar glass plates ($D^*$) were a factor of ~ 2.5 larger than those in bulk, which was observed previously by us for EAN.[14] Just after placement in the magnetic field, the diffusivity of the cations gradually decreased, as shown in Figure 2. The process reaches saturation after ~ 10 h or longer with a final equilibrium value of the diffusion coefficient, $D^*_{MF}$. It was also shown for EAN confined between polar glass plates that after removing the sample from the magnetic field, $D$ reversibly changed from $D^*_{MF}$ to $D^*$. The same trend was observed for the sample of confined PAN.

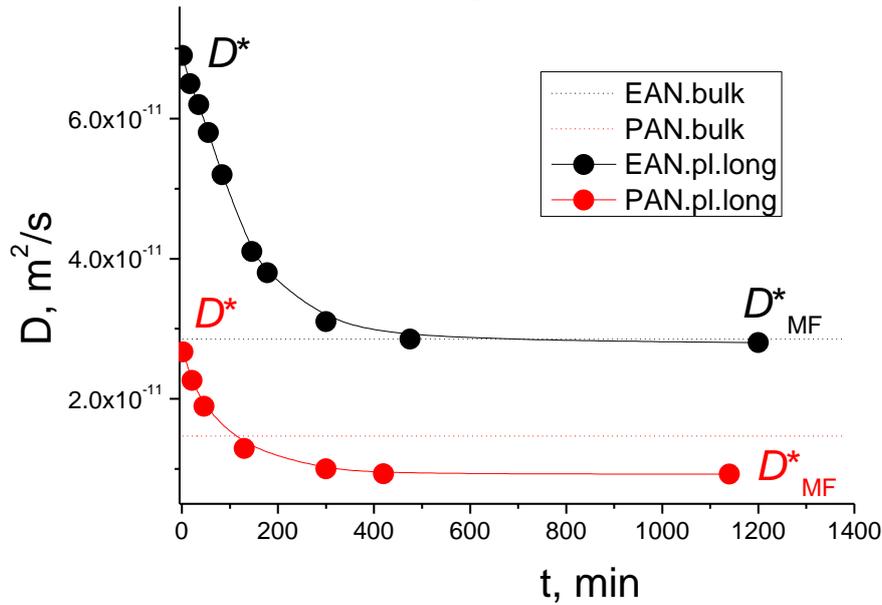

**Figure 2.** Change in diffusion coefficients of EAN (black) and PAN (red) cations after placing ammonium nitrate ILs confined between polar plates in a magnetic field of 9.4 T. Diffusion coefficients were measured along the plates, $t_d$ = 3 ms. Dotted lines correspond to bulk ILs.

Diffusion coefficients of PAN in both bulk and confinement are a factor of 2-3 lesser than those of EAN. This might be related to the larger size of the propylammonium cation and also to the possible difference in glass transition temperatures of EAN and PAN. However, the forms of the curves $D(t)$ for these two ILs are qualitatively similar (see Figure 2). Hence, the translational dynamics of ions in the studied ammonium nitrates between polar glass plates were different from those in the bulk phase $I_B$. Furthermore, these ILs confined between polar glass plates formed two types of equilibrium dynamic phases: the "$I$" phase in the absence of the magnetic field and the "$I_{MF}$" phase, which forms after exposure to the magnetic field. Both of these phases are dynamically isotropic, because $D$ is the same along and perpendicular to the static magnetic field.[13,14]

Some of characteristics of these phases in EAN were reported earlier.[14] It was shown that the kinetics of $I \rightarrow I_{MF}$ transformation did not depend on the sample orientation and demonstrated coincident curves for plates oriented along and perpendicular to the magnetic field.[14] We also performed an additional experiment described below: The sample with plates oriented parallel to the magnetic field was first exposed to the field until the cation diffusion coefficient reached $D_{MF}*$. Immediately afterwards it was turned 90°, so the plates were oriented perpendicular to the field. Measurements showed the same value of the diffusion coefficient, $D_{MF}*$. This experiment additionally suggested that the orientation of the plates relative to the static magnetic field had no effect on the translational dynamics in the **I$_{MF}$** phase.

### 3.2. Kinetics of phase transformations

The change in the phase characteristic over time (a kinetics of a process) occurring at constant temperature can be analyzed by a heuristic Avrami equation[18,19] (also known as a Johnson-Mehl-Avrami-Kolmogorov (JMAK) equation). This equation describes processes of the autocatalytic type, for example a phase transformation occurring by a nucleation mechanism. The equation can specifically describe the kinetics of crystallization or other changes of phases in a material. The degree of phase transformation (fraction of a growing phase) is described as follows:[18,19]

$$Y(t) = 1 - \exp(-Kt^n), \qquad (4)$$

where $K$ is a reaction rate constant and the exponent $n$ is a constant related to the nucleation and growth.

In our experiments the parameter characterizing the phases is the diffusion coefficient. Therefore, the degree of phase transformation (fraction of phase **I$_{MF}$** formed in ILs between polar glass plates under the influence of the magnetic field) can be calculated as in Eq. (5):

$$Y(t) = (D_{init} - D(t))/(D_{init} - D_{fin}), \qquad (5)$$

where $D_{init}$ is $D*$, and $D_{fin}$ is $D*_{MF}$ (see Figure 2). The calculated $Y(t)$ for the kinetics of transformation of EAN and PAN are shown in Figure 3 (circles). Dependences, $Y(t)$, were fitted by Eq. (4) by the least square method (lines in Figure 3). Avrami parameters $K$ and $n$ used for the fittings are also reported in Figure 3.

For both EAN and PAN, the Avrami exponent $n$ is close to 1. Originally, $n$ was held to have an integer value between 1 and 4, which reflected the nature of the transformation.[20] There were simulations to relate dimensionality and mechanism of crystallization with the observed Avrami exponent $n$. In earlier work of Cahn, $n = 1$ was related to a case of site saturation at nucleation on a surface.[20] In a more recent work of Yang et al.,[21] it has been shown that homogeneous as well as heterogeneous nucleation in the surface-independent and diffusion-independent deposition rates gives $n \sim 1$ in a 3D system. This suggests the formation of a primary nucleus of a new phase spontaneously or on heterogeneous nucleation agents in the beginning of the process and further growth of the new phase by deposition of molecules from the original phase on the nucleus surface. The rate constant $K$ is influenced by a ratio of interfacial to thermal energy. In our data, $K$ is higher for PAN in comparison with EAN (see Figure 3), because its thermal energy per mass of the PA cation is lower.

Inspecting fits presented in Figure 3 one can conclude that the kinetics of transformations of phases in EAN and PAN exposed to a magnetic field can be described satisfactorily well by the Avrami equation using parameters characteristic for autocatalytic processes, which often are controlled by spontaneous nucleation.

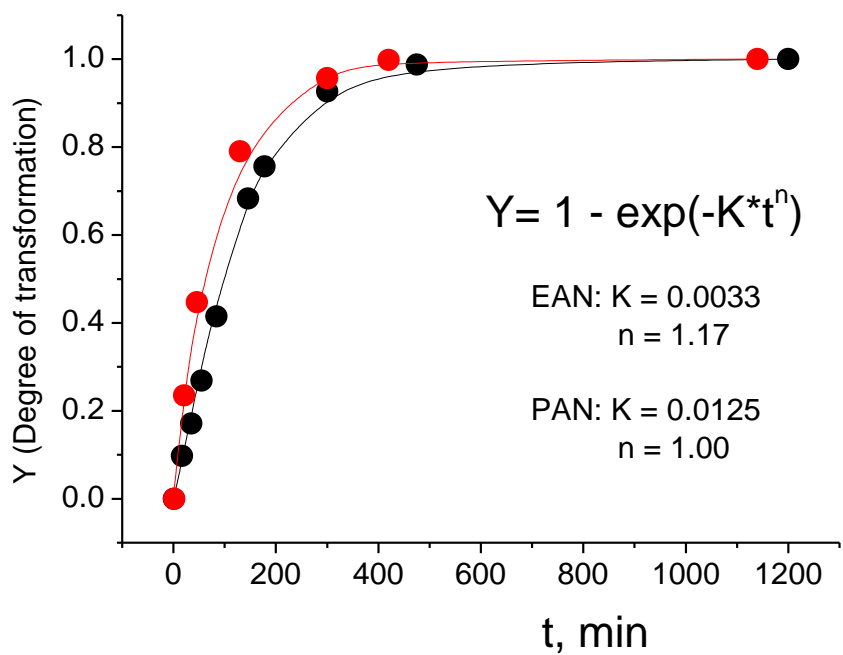

**Figure 3.** Degree of phase transformation (fraction of phase $I_{MF}$) as a function of time for EAN (black) and PAN (red) confined between polar glass plates. *Y* was calculated using Eq.(5) and experimental data were taken from Figure 2. Symbols are experimental points, while lines are best fits using the Avrami equation (Eq. (4)) with indicated values *K* and *n*.

### 3.3. Freezing and melting of EAN confined between polar glass plates

Freezing and melting of bulk EAN have been studied by differential scanning calorimetry technique by Salgado *et al*.[22] The latter data clearly revealed melting and solidification peaks, thus confirming that EAN in bulk is a very good crystal-forming material. However, thermodynamic characteristics of the sample with EAN confined between polar glass plates used in our experiments are difficult to study by the standard calorimetry technique, because the sample present a solid block containing mostly glass (~97% by volume), while thin IL layers between glass plates are quite thermally isolated by the glass plates. Nonetheless, we can observe the phase transition indirectly by NMR, because NMR relaxation times, especially the NMR transverse relaxation time $T_2$, undergo a rapid change at the liquid-to-solid and solid-to-liquid transitions.[23] We studied liquid-solid and solid-liquid phase transitions in the same sample prepared as **I** and $I_{MF}$ phases (the sample was either not exposed or exposed for 24 h to the magnetic field) in two experiments. In the first experiment, the sample was placed in the probe of NMR spectrometer, after which the temperature was steadily decreased from 293 K to 256 K. The cooling rate of 1 K·min$^{-1}$ was chosen to minimize the effect of $I \rightarrow I_{MF}$ transformation in the course of the experiment for the sample in **I** phase and to maintain similar experimental conditions for the sample in $I_{MF}$ phase. Integrals of the $^1$H NMR spectra were obtained from Free Induction Decays (FID) and plotted as a function of temperature. Results of these experiments are shown in Figure 4A.

Figure 4A shows that the amplitude of the signal did not change down to a certain temperature at which it abruptly dropped down. An abrupt decrease of the signal of **I** phase occurs at ~ 263 K and the signal of $I_{MF}$ phase at ~ 266 K. These temperatures are much lower than that of bulk EAN (~285 K).[24] In addition, there were differences in the solidification behaviour of **I** and $I_{MF}$ phases. For $I_{MF}$ the signal completely disappeared at temperatures below 266 K (red line in Figure 4A). However, for the **I** phase (black line in Figure 4A), typical liquid $^1$H NMR spectra of EAN with an integral amplitude of ~ 0.24 was maintained down to 256 K, the lowest temperature of this experiment.

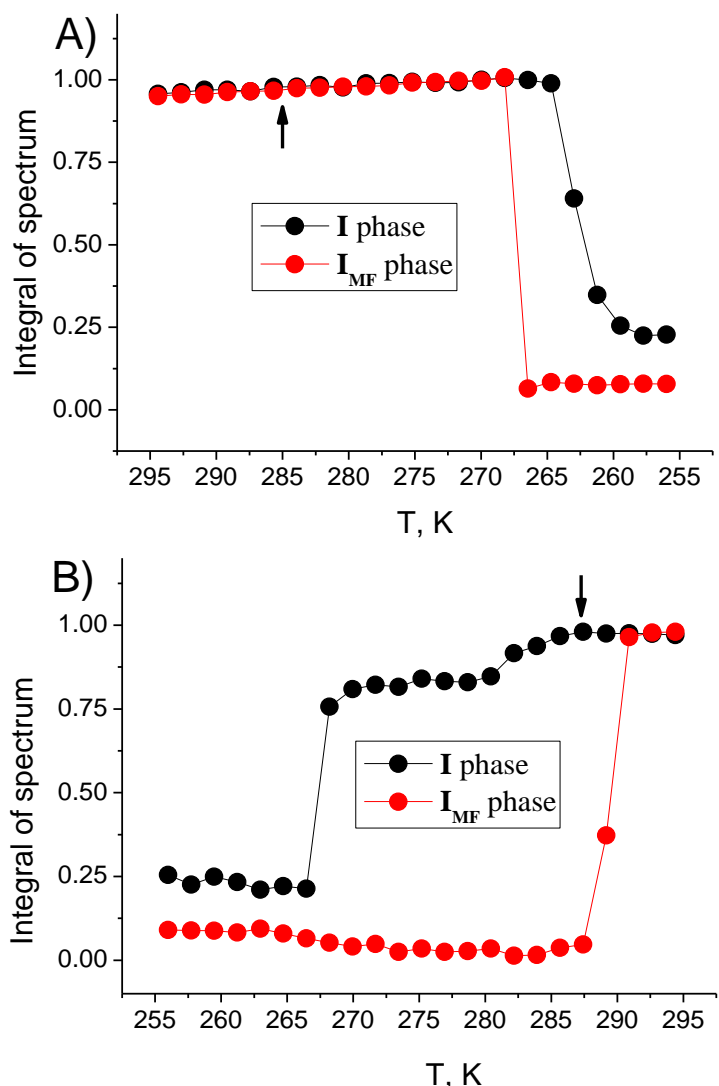

**Figure 4.** Integral intensities of $^1$H NMR spectra of EAN confined between polar glass plates before (phase **I**) and after (phase $I_{MF}$) exposure of the sample in the magnetic field as a function of temperature at **A**) cooling (rate 1 K·min$^{-1}$) and **B**) heating (controlled heating rate). Arrows show the melting temperature of bulk (unconfined) EAN.[24]

Freezing of EAN in our experiments was performed under non-equilibrium conditions, because of a rather high cooling rate (1 K·min$^{-1}$). Additionally, the process of crystallization is triggered by a nucleation mechanism and it takes place in an overcooled liquid, i.e. essentially not at equilibrium. In our second experiment, we studied melting of EAN, a process that can be led at conditions much closer to an equilibrium. The sample of confined EAN prepared in **I** or **I$_{MF}$** phases was cooled down to 256 K, maintained at this temperature for a few hours and then slowly heated stepwise (with a step of 1 K) under controlled heating rate. After each heating step, a series of $^1$H NMR spectra were obtained with an interval of *ca.* five minutes until intensities of $^1$H NMR resonance lines did not change further. Then the $^1$H NMR spectrum was stored and the sample was heated again by 1 K and so on. The resulting heating curves for the sample in both **I** and **I$_{MF}$** phases at such 'quasiequilibrium' melting conditions are shown in Figure 4B. Melting of the **I$_{MF}$** phase occurred sharply at 289 K. For the **I** phase, the process of melting began at lower temperatures, *ca.* 270 K, and "smeared" over the broad temperature range up to ~285 K.

$^1$H NMR spectra obtained for phases **I** and **I$_{MF}$** at melting are shown in Figures 5A and 6B, respectively. For the **I** phase (Figure 5A), the sample at heating demonstrated spectral lines of protons in all three chemical groups of the ethylammonium cation and also a signal from water at ~ 4.7 ppm. In particular at lower temperatures, the contribution of the water NMR signal in the spectra of the **I** phase is much larger than the real fraction of water protons in the sample, because of the much longer NMR relaxation times of H$_2$O protons in comparison with protons of EAN and also because of an increase of the $^1$H NMR signal at lower temperatures (Curie's law[25]). Heating the sample up to 270 K did not significantly change the spectral form apart from minor changes in chemical shifts for NH$_3^+$ protons (from *ca.* 7.5 to 7.6 ppm) and for H$_2$O protons (from *ca.* 4.9 to 4.7 ppm) in the temperature interval from 256 to 266 K, respectively. In the temperature range of 270 – 285 K, spectral line intensities of ethylammonium progressively increased, while the contribution of the water signal decreased. Finally, the water signal almost disappeared at 293 K.

For the **I$_{MF}$** phase (see Figure 5B), the sample at heating in the range of 256-288 K did show only very weak spectral lines. This means that at these temperatures EAN was almost completely 'frozen' solid material without any significant motion of ions. In solids both the dipole-dipole interaction between protons and the $^1$H chemical shift anisotropy, which are 2$^{nd}$ rank interactions with a strong orientation dependence, have dramatically broadened $^1$H NMR resonance lines. The $^1$H NMR spectral lines appeared at T > 289 K and the spectral shape was maintained at higher temperatures, demonstrating that EAN in this sample sharply and completely melted at 289 K.

No any signal from water was observed in $^1$H NMR spectra for the **I$_{MF}$** phase. This may occur if water protons are at conditions of "fast exchange" with protons of –NH$_3$ groups of EAN.[23] Usually the "fast exchange" of protons of two chemical groups, which differ in chemical shift leads to averaging of chemical shifts and producing a line in the NMR spectrum with an intermediate value of chemical shift. Because -NH$_3$ protons in our system are much in excess, the exchange leads to disappearance of NMR spectra line characteristic to bulk water for the **I$_{MF}$** phase. This explanation also suggests that water protons are at condition of "slow exchange" in the **I** phase. Thus, **I** → **I$_{MF}$** transformation leads to change of dynamic state of water protons in the studied system from "slow exchange" to "fast exchange".

An alternative explanation of the disappearance of the water signal in NMR spectrum of the **I$_{MF}$** phase is the increase of NMR relaxation rate of the water protons. The relaxation rate became so short that water protons did not contribute to the $^1$H NMR echo signal. Such short $^1$H NMR relaxation of water protons may be due to the fact that water is expelled from EAN to the glass surfaces. It is known that adsorbed water possess 'solid-like' properties[26] and, therefore, it does not contribute to $^1$H NMR echo signal of a highly mobile liquid phase.

Therefore, it can be concluded that the melting behaviour of the **I** and **I_MF** phases is significantly different. **I** phase melting proceeded in a rather broad temperature range, while in the **I_MF** phase the complete melting occurred in a temperature range of only *ca.* 1 K. Additionally, the **I_MF** phase did not show any residual water NMR signal.

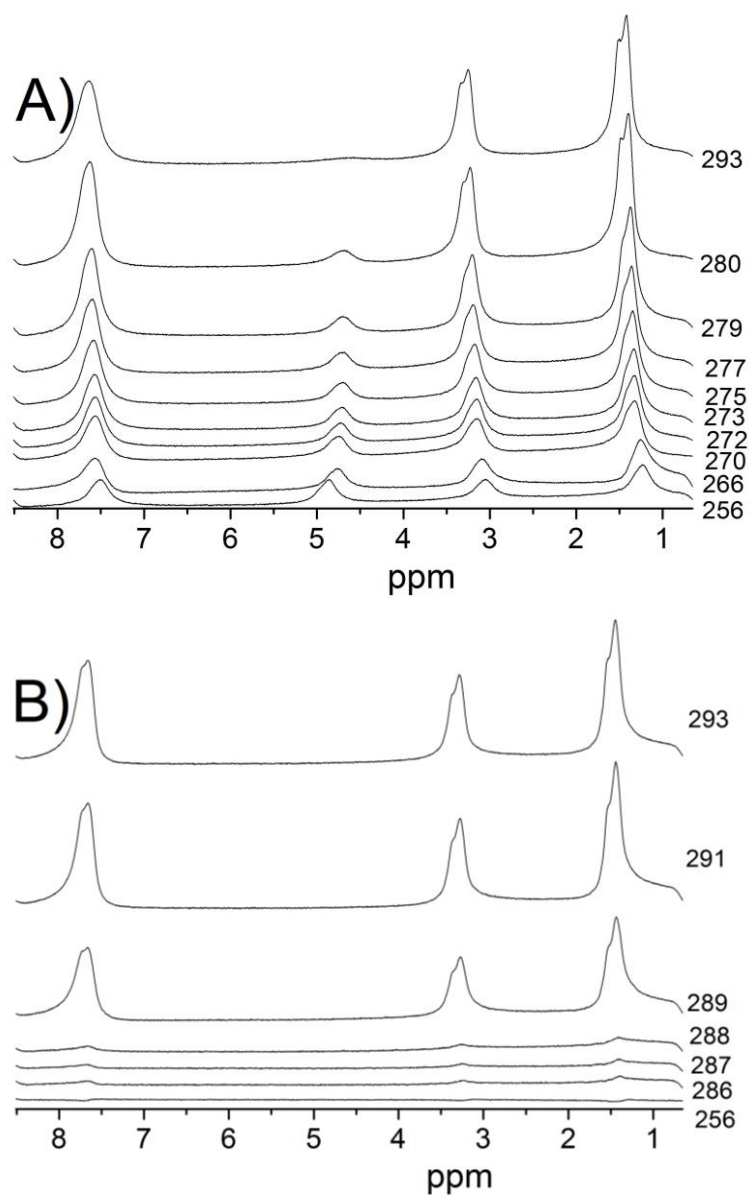

**Figure 5.** $^1$H NMR spectra of EAN confined between polar glass plates: **A)** before (phase **I**) and **B)** after (phase **I_MF**) exposure to the magnetic field as a function of temperature during heating from 256 to 293 K.

It has been reported[14] that the reverse transformation of **I_MF** → **I** occurring after removing the sample from the magnetic field has taken a few days at room temperature. We performed a separate experiment, in which the sample after a long-term exposure to the magnetic field in the NMR spectrometer (**I_MF** phase, demonstrating $D^*_{MF}$) was placed in a freezer at 193 K (-80 °C) and kept there for 12 days. Afterwards, the sample was placed back in the NMR spectrometer at room temperature (*ca.* 293 K) and diffusion of EAN was then measured after the sample has completely thawed and the system has reached a new thermal equilibrium. The measurement showed that the diffusion coefficient of EAN was equal to $D^*_{MF}$, thereby demonstrating that the transformation of $\mathbf{I_{MF}} \rightarrow \mathbf{I}$ was stopped in the frozen state of EAN.

### 3.4. Phases formed in EAN confined between polar glass plates under the influence of an external magnetic field

Recently, we reported on the reversible alteration of diffusivity of ions and NMR relaxation rates of protons in ethylammonium nitrate confined between polar glass plates, when the sample was exposed to a permanent magnetic field of 9.4 T during a few hours and then taken away from the magnetic field.[14] In this our study we further explored another protic ionic liquid of the same class, propylammonium nitrate, and found that this IL also demonstrated a similar response on the external magnetic field. We suggested that the process occurring in the magnetic field can be related to a phase transformation. In this study, we showed that the transformation process needs molecular mobility; we further investigated freezing and melting behavior of the sample with EAN confined between polar glass plates, depending on prehistory.

The kinetics of phase transitions of confined EAN and PAN well fits with Avrami phase transition theory. This demonstrates that the process is autocatalytic and very likely proceeds through a nucleation mechanism. The obtained value of the Avrami equation exponent *n* is close to 1, which can be described by homogeneous or heterogeneous nucleation in the surface-independent and diffusion-independent deposition rates in a 3D system.

Temperature measurements of solidification and melting behaviors showed a difference in phase transition temperatures, interval of the phase transition temperatures and completeness of these transitions for confined phases of EAN formed before and after exposure in the magnetic field. An unexpected result was obtained that concerns the state of residual water under transformation of confined EAN under exposure to a magnetic field.

Residual water, which is often present in EAN, could modify studied system. EAN/$H_2O$ and some other protic IL aqueous bulk mixtures have been studied earlier.[27-32] Simulations showed that water molecules at low $H_2O$ concentration in the IL do not destroy local sponge-like structure specific to the bulk neat EAN,[30,31] instead they accommodate themselves in the network of hydrogen bonds[31] of the IL or leads to swelling of the polar regions.[30] According to our data for the sample of EAN confined between polar glass plates, water protons, probably, can change their exchange conditions with –$NH_3$ group of EAN from "slow exchange" to "fast exchange".

In this study we confirmed our previous suggestion that the alteration in the dynamic properties of confined EAN subjected to a magnetic field is related to the formation of real physical-chemical phases. However, we did not identify the exact molecular mechanisms leading to formation of confined dynamic phases of EAN in the presence and absence of a magnetic field. Further, experimental studies and simulations intended to reveal structural and dynamic properties of these phases are being undertaken in our laboratory.

## 4. Conclusions

We demonstrated that different dynamic phases found earlier in ethylammonium nitrate confined between polar glass plates can be found in other ammonium nitrate ionic liquids. The process of altered dynamics of ionic liquids confined between polar glass plates and placed in a strong magnetic field can be described well by the Avrami equation, which is typical for autocatalytic (particularly, nucleation controlled) processes. The transition can be stopped by freezing the sample. Cooling and heating investigations showed differences in freezing and melting behavior of the sample that had been exposed or not exposed to the magnetic field. The sample after exposure to the magnetic field demonstrated no signal from residual water in the $^1$H NMR spectra. Generally, our findings confirm our suggestion that the alteration in dynamic properties of confined ethylammonium nitrate ionic liquids exposed to a magnetic field is related to the formation of real physical-chemical phases.

Given that properties of structure and dynamics of ionic liquids are significantly altered in confinement and in the magnetic field, we suggest that the results will have strong implications for interface-intensive applications of ILs, such as their use in machinery, electrochemical and electro-magnetic systems.

**Acknowledgments**

Dr. Sh. Bhattacharyya is acknowledged for his supervising the ethylammoniun nitrate and propylammoniun nitrate syntheses, and for sample characterization. "Scriptia Academic Editing" is acknowledged for English correction and proof-reading of this manuscript.